# Enhancing Co-packaging Optics Enabled Silicon Photonics Security Assurance Hardware Fingerprinting

Liton Kumar Biswas, M Shafkat M Khan, Himanandhan Reddy Kottur,
Hao Wang, Hamed Dalir, and Navid Asadizanjani
Department of Electrical and Computer Engineering, University of Florida
5000 Malachowsky Hall, 1889 Museum Road
Gainesville, Florida-32611, USA
Email: litonkumarbiswas@ufl.edu

## Abstract

Silicon photonics (SiPh) enables integration of optical components using standard semiconductor processes, greatly improving data communication bandwidth and energy efficiency. However, photonic integrated circuits (PICs) face unique security challenges, such as counterfeit or tampering threats, that conventional electronic security methods do not address. We propose a novel hardware fingerprinting technique that embeds two-dimensional photonic crystal (PhC) patterns into the density-control filler regions of a PIC. Each PhC pattern is designed to resonate at specific visible to near-infrared wavelengths, producing a distinctive optical signature (based on wavelength, polarization, and incident angle) for each device. Finite-difference time-domain (FDTD) simulation using ANSYS Lumerical is employed to optimize nanostructure dimensions and spacing so that each device's reflection/absorption spectrum contains unique narrowband peaks. No extra fabrication steps or materials are required beyond standard lithography, keeping costs low. The embedded nanostructures have sub-50 nm precision, making forgery extremely difficult. Our method yields a high-resolution, scalable fingerprint for silicon photonic chips, enabling cost-effective device authentication and improved supply chain security.

## I. Introduction

Silicon photonics (SiPh) has emerged as a key technology to meet increasing demands for higher bandwidth, lower latency, and improved power efficiency [1]. By leveraging silicon to integrate optical components, SiPh significantly enhances high-speed data communication and large-scale integration while using existing semiconductor fabrication processes to reduce cost [2]. The growth of applications such as artificial intelligence (AI), cloud computing, and the Internet of Things (IoT) has further driven SiPh adoption into critical sectors like medical devices, defense systems, and aerospace [3].

However, the increasing reliance on photonic integrated circuits (PICs) for critical applications introduces new security and reliability challenges [4]. PICs are highly sensitive to physical or cyber tampering, and introducing counterfeit components can easily degrade their performance or cause failures [5]. Traditional security measures designed for electronic circuits often fall short in addressing these unique photonic threats [6]. This growing concern has motivated research into novel security techniques for SiPh devices.

Hardware fingerprinting is a promising approach to securing SiPh systems. It creates unique, device-specific signatures by integrating identifiable patterns into the chip during fabrication [7]. In contrast to conventional methods that require extra fabrication steps or special materials, our technique repurposes the filler patterns commonly used in PIC manufacturing as unique signature patterns. These filler patterns, which traditionally ensure uniform lithography density, are instead designed as two-dimensional photonic crystal structures. Each embedded PhC is engineered to resonate at specific wavelengths in the visible to near-infrared (NIR) range, yielding a distinct optical signature that can be read with relatively low-cost equipment. The resonant nanostructures are specified with sub-50 nm precision, ensuring each device's fingerprint is unique and highly resistant to cloning.

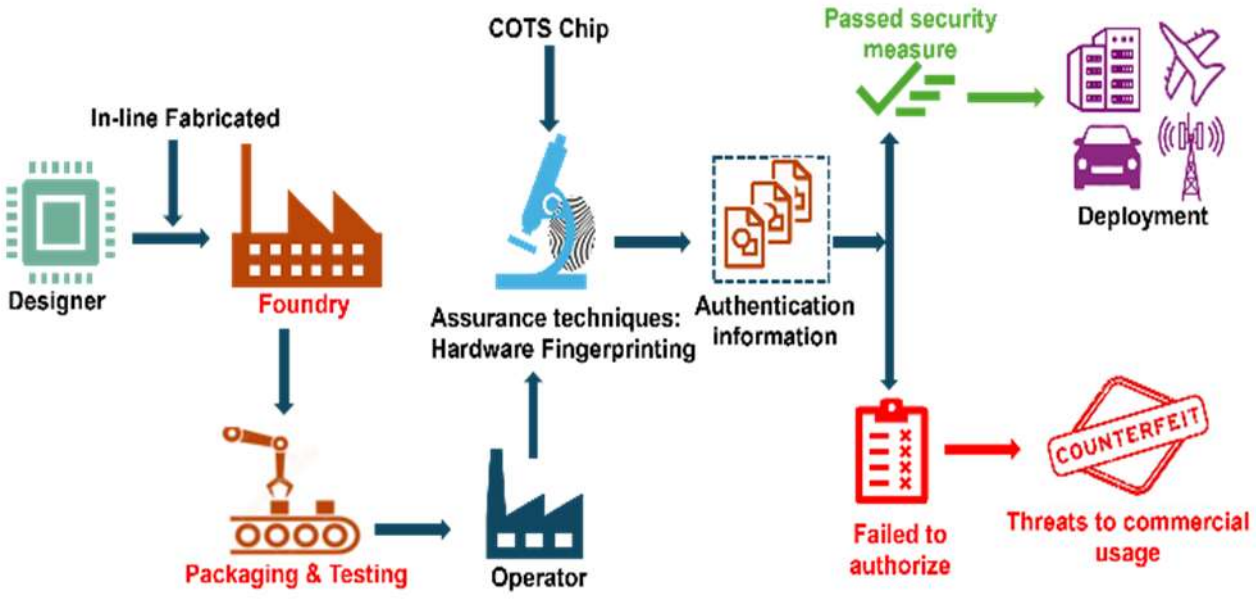

*Fig. 1: SiPH supply chain vulnerability and assurance technique to combat threats*

In this paper, we propose and evaluate this novel PhC-based fingerprinting method. We design and simulate silicon nanostructures that encode information via their intrinsic properties (wavelength, polarization, incident angle). Using ANSYS Lumerical FDTD simulation, we optimize the structures' size, shape, and spacing so that each pattern produces unique reflection and absorption peaks (essentially distinct color outputs). The combination of controlled design and inherent process variability enables robust device authentication, strengthening the overall security of silicon photonic systems.

## II. Background

### *A. Antenna and Pillar Structures*

In silicon photonics, optical antennas and pillar structures are nanofabricated elements that manipulate light on chip. Pillar arrays form two-dimensional photonic crystal lattices that can trap or filter light at specific wavelengths. These periodic pillar structures (or their complementary holes in a silicon slab) act as optical resonators, creating sharp spectral features. Similarly, optical antenna structures (analogous to radio antennas) include grating couplers or metasurface patches used to efficiently couple light into or out of waveguides. The exact geometry of these dielectric nanostructures—such as the pillar diameter, height, and lattice spacing—strongly influences their optical properties [8]. For example, varying the size or spacing shifts the resonance wavelength and changes the absorption/reflection spectrum. In our design, we exploit such pillar-based photonic crystals as the basis for encoding unique device fingerprints.

### *B. Density Control as a Polarization-sensitive Pattern*

Density control patterns [9] are specially designed filler layouts used in PIC fabrication to maintain uniform material density across the chip. Uniform density is critical for reliable lithography and etching: variations can lead to uneven exposure or etching depth across the wafer. By carefully arranging pillars or holes, the overall pattern density remains constant even as functional circuits are distributed. In this work, these density-control fillers are repurposed to carry the hardware fingerprint. The spatial density and arrangement of the photonic crystal nanostructures are chosen not only to meet fabrication uniformity requirements, but also to encode information through their optical response. Photonic crystal structures can be polarization-sensitive [10], meaning their optical response depends on the polarization state of incident light. For instance, an array of asymmetrical pillars or a birefringent material can reflect or transmit TE- and TM-polarized light differently. We exploit this effect to add another encoding dimension. By designing nanostructures that support polarization-dependent resonances, the device fingerprint can carry information in how the response varies with polarization. In practice, the fingerprint reading system may illuminate the pattern with different polarization states and observe changes in reflected color or intensity. This polarization sensitivity, combined with wavelength-selective resonance, provides multiple "layers" of encoding, further increasing the difficulty of replication. The overall fabrication process thus remains unchanged, while the embedded PhCs simultaneously serve as unique signature generators.

### *C. Fabrication Process Variations*

Even with advanced lithography, fabrication of nanostructures exhibits inherent variations at the nanometer scale. Fabrication process variations include small deviations in feature size, shape, or placement caused by factors like resist exposure nonuniformity or etch fluctuations. Our fingerprinting method leverages these unavoidable variations: the photonic crystal patterns are designed with target dimensions and sub-50 nm tolerances, so that any slight deviation translates into a measurable difference in the optical signature [11]. For example, a tiny change in a pillar diameter shifts the resonance wavelength. The high-resolution nature of the design makes even minor fabrication errors detectable. In effect, the natural process variation becomes part of the device's unique signature, and counterfeiting (which would likely introduce larger errors) becomes immediately apparent.

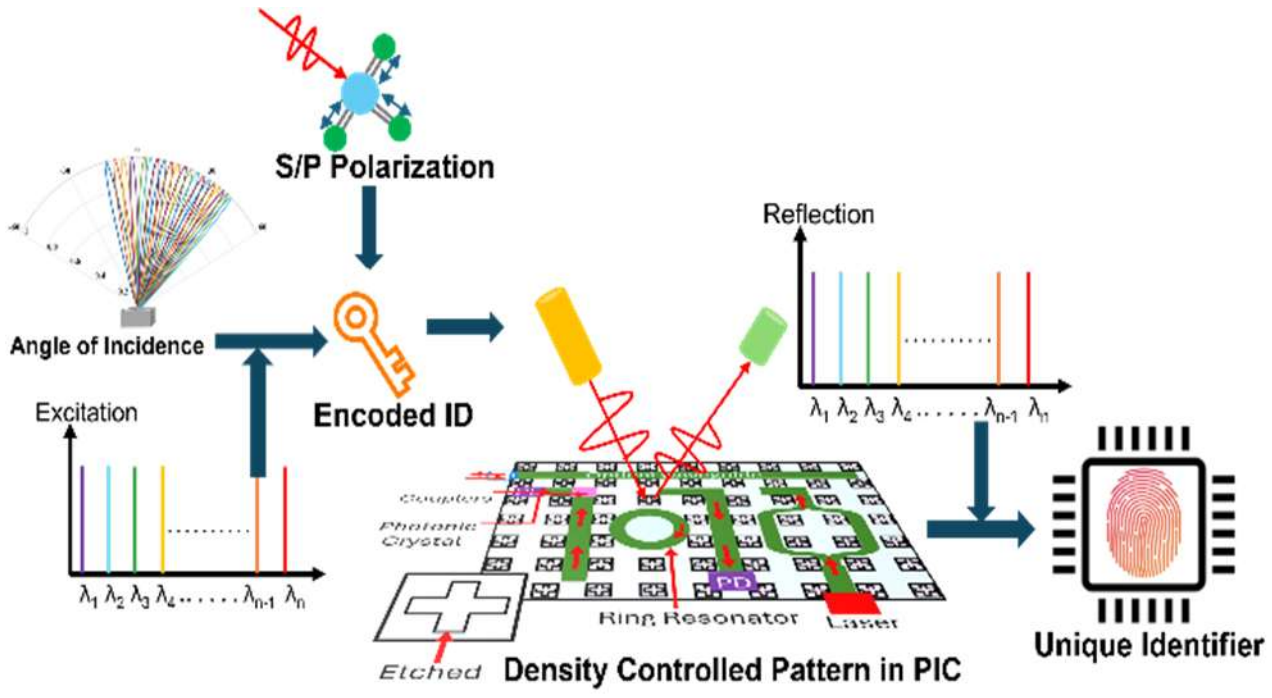

*Fig. 2: A framework representation of hardware fingerprinting*

### D. *Hardware Fingerprinting*

Hardware fingerprinting refers to the creation of unique, unclonable identifiers embedded in a physical device [12]. In silicon photonics, this means integrating distinct nanostructure patterns during fabrication so that each PIC has its own "fingerprint." Because we use standardized design patterns and rely on process variation, no extra materials or steps are needed. Each device's fingerprint emerges naturally from its manufactured geometry. At any point in the supply chain, an optical measurement of the fingerprint (e.g., by scanning light and detecting its spectral response) can verify authenticity. This approach provides a robust defense against tampering or counterfeit components, since an adversary would need to exactly reproduce the tiny nanostructure details across the entire pattern to fool the system. As described above, the combination of precise design and process variation ensures that each device's fingerprint is inherently unique and verifiable.

## III. Design and Simulation Concept

### A. *Silicon Photonics Supply Chain Vulnerability and Assurance*

The growing reliance on photonic integrated circuits (PICs) for mission-critical applications—from AI accelerators to aerospace and defense—demands an equally robust system for device verification and supply chain security. In traditional electronics, techniques such as PUFs (Physical Unclonable Functions), electrical signature-based IDs, and cryptographic authentication have helped mitigate counterfeit and tampering risks. However, photonic circuits introduce a different set of challenges. Their passive and active optical components are fabricated using nanometer-scale precision, and even small perturbations in geometry can drastically affect performance. Yet, these same sensitivities also open a novel opportunity: using nanostructure-level optical phenomena as identifiers.

Silicon photonic devices are typically manufactured through foundry services, and pass through several stages—design, tape-out, wafer fabrication, die singulation, assembly, packaging, and test—before reaching the end-user. At any of these stages, an adversary could potentially introduce counterfeit components, make unauthorized modifications, or swap legitimate devices. Current anti-counterfeiting techniques, such as serialization, external labels, or package-based security, do not provide sufficient protection against physical-level tampering within PICs. Moreover, optical layers in photonic chips are difficult to probe nondestructively once packaged.

Hardware fingerprinting through photonic crystal (PhC) nanostructures embedded in density-controlled regions offers a solution that is intrinsic, unclonable, and fabrication-integrated. It enables secure device authentication without needing any additional circuitry or active device modification. Since each fingerprint is encoded optically and embedded at the chip level during fabrication, the device can be verified using external, nondestructive optical interrogation. This significantly reduces inspection overhead while improving confidence in component provenance. Authentication can be performed at multiple checkpoints in the supply chain using simple reflectometry or hyperspectral imaging tools.

### B. *A framework representation of hardware fingerprinting and schematic of Si-based nanostructures*

The proposed fingerprinting scheme is inspired by traditional anti-counterfeiting holographic stickers, which are widely used in product packaging and financial documents. These holograms exploit nanoscale surface structures to diffract light and produce angle- and polarization-dependent visual effects—iridescent colors, rainbow patterns, and depth illusions. However, those holograms are fabricated using metal films, typically aluminum, and require complex multi-step lithographic stamping processes or embossing. They are surface-attached, not embedded in the functional substrate, and are often vulnerable to environmental degradation and physical removal.

In contrast, our method embeds the fingerprint directly within the silicon layer of a photonic chip, leveraging dielectric materials and lithography processes already used in PIC fabrication. The security feature is thus both functional and structural—it cannot be removed without damaging the device. Unlike metal-based holograms, which rely on plasmonic effects or surface diffraction, our system uses high-index dielectric photonic crystals to induce resonance-based spectral filtering. This approach provides superior optical stability, durability, and compatibility with CMOS-based silicon photonics.

The framework comprises four main stages:

*a) Design of Photonic Crystal Patterns:*

The fingerprint area consists of a periodic array of subwavelength silicon pillars or holes in a $SiO_2$ substrate. The PhC lattice constants, fill factors, and thicknesses are selected so that the structure supports guided-mode resonance at specific wavelengths within the visible to near-infrared (NIR) spectrum. Each variation in geometry (e.g., pillar diameter, gap, height) produces a distinct spectral signature. These spectral features act like a "color barcode" unique to each chip.

*b) Simulation Using FDTD Methods:*

Full-wave 3D simulations are performed using the ANSYS Lumerical Finite-Difference Time-Domain (FDTD) solver. These simulations capture how incoming broadband light interacts with the nanostructures and reveal the reflected and absorbed power spectra. By sweeping design parameters, we identify configurations that produce sharp, distinguishable resonance peaks with full-width half-maximum (FWHM) of less than 10 nm. The simulations also account for realistic fabrication variations and angle/polarization dependence, which are incorporated to further diversify the encoding space.

c) *Fabrication with Standard PIC Processes:*

The photonic crystal fingerprints are integrated into filler regions—areas of the chip layout typically reserved for density uniformity. These regions do not interfere with optical circuits and allow passive security embedding. No extra process steps are required; e-beam or stepper lithography and plasma etching create the fingerprint patterns alongside the PIC features. Unlike conventional security tags that must be applied post-fabrication, this method is inherently tamper-resistant and built into the chip's physical layer.

d) *Detection and Authentication:*

The fingerprint is interrogated using polarized, angled light or a hyperspectral scanner. Reflected light is measured and matched against a secure fingerprint database. Because the encoded information depends not only on wavelength but also on incident angle and polarization, a rich multidimensional signature is generated. The uniqueness of each pattern is due to both intentional design and natural nanofabrication imperfections, which add entropy to the system.

Notably, this approach requires no active powering or signal routing to the fingerprinted region, unlike many electrical PUFs or embedded ID schemes. It is optically readable, passive, scalable, and compatible with wafer-level or packaged-level inspection.

In summary, the use of silicon dielectric photonic crystals for hardware fingerprinting provides a durable, high-resolution, and fabrication-compatible solution for ensuring authenticity in SiPh systems. It extends the concept of holographic verification into the realm of nanophotonic security, with advantages in resilience, spectral precision, and seamless integration into semiconductor production flows.

## IV. Result Interpretation on Test Vehicle

### A. Simulation Result Analysis

To validate the proposed fingerprinting mechanism and evaluate its effectiveness, a comprehensive simulation-based investigation was conducted using the Finite-Difference Time-Domain (FDTD) solver from ANSYS Lumerical. The simulations focused on photonic crystal (PhC) patterns comprising silicon pillars arranged in a square or hexagonal lattice within a silicon dioxide ($SiO_2$) background. The periodicity, pillar diameter, height, and inter-pillar spacing were systematically varied to investigate their influence on optical resonance properties.

The simulations revealed that the PhC nanostructures exhibit distinct resonance behavior in the visible to near-infrared (NIR) spectral range (400–1000 nm). These resonances are characterized by strong peaks in the reflectance and absorption spectra at specific wavelengths. Each variation in geometrical parameters shifts the resonance peak, thereby producing a unique spectral signature. For example, increasing the pillar diameter from 150 nm to 220 nm in steps of 10 nm results in a redshift of the primary resonance peak by approximately 30–50 nm. Similarly, decreasing the inter-pillar gap leads to stronger field confinement and narrower full-width at half-maximum (FWHM), enhancing spectral resolution.

In addition to resonance wavelength shifts, the polarization and angle sensitivity of the structures was also studied. Structures with broken symmetry—such as elliptical or rectangular pillars—exhibited polarization-dependent behavior. The reflectance spectra for transverse electric (TE) and transverse magnetic (TM) polarizations showed noticeable variation, confirming that polarization can be leveraged as an additional encoding parameter in the fingerprinting scheme.

Furthermore, the angular dependence of the resonance peaks was investigated by simulating incident angles from 0° (normal incidence) to 30°. As expected from guided-mode resonance theory, higher incident angles shifted the resonance peaks slightly and modulated their intensity. These effects can be incorporated into a multidimensional fingerprint space, enabling more robust authentication.

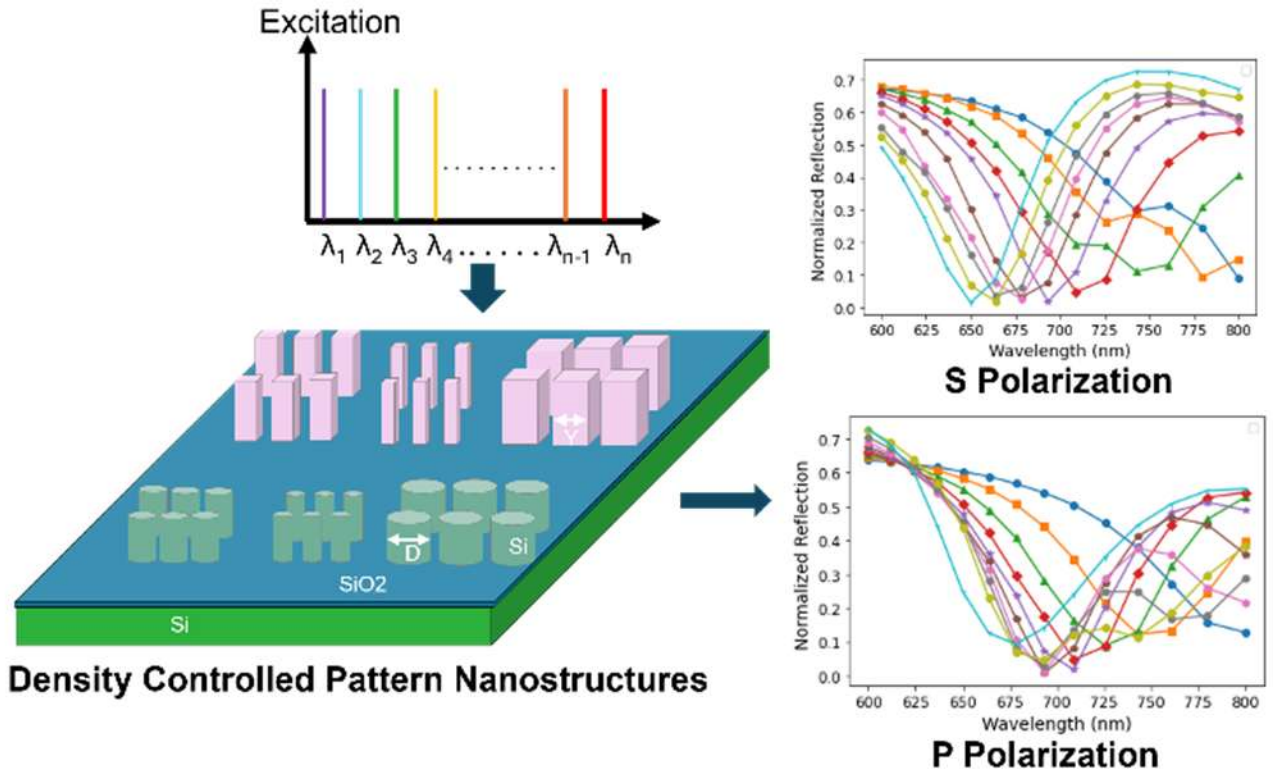

*Fig. 3: Schematic of Si-based nanostructures with different dimensions reflection spectra under s-polarization and P-polarization*

To assess the practicality of using these spectral features as fingerprints, simulations were carried out for over 100 variations in pattern geometry. The resulting spectra were compared to evaluate uniqueness and separability. The minimum separation between any two resonance peaks was found to be greater than 20 nm, and in most cases, over 40 nm. This ensures that the patterns are sufficiently distinct to avoid false positives in an authentication system.

Additionally, FWHM values for resonance peaks were consistently below 15 nm, offering high selectivity.
By compiling the simulated spectral responses into a fingerprint database, we demonstrated the feasibility of mapping each nanostructure design to a unique identifier. These identifiers can be read optically with standard spectrometers or hyperspectral cameras. The simulations conclusively showed that both design-controlled and process-variation-enhanced signatures are achievable, forming the basis for secure and scalable silicon photonic hardware authentication.

### *B. Fabrication Imaging Analysis*

To further validate the simulation results and assess the feasibility of real-world implementation, photonic crystal fingerprint structures were fabricated on silicon-on-insulator (SOI) wafers using standard PIC lithography and etching processes. The fabrication flow included electron-beam lithography for pattern definition, reactive ion etching (RIE) for vertical profile formation, and a buffered oxide etch (BOE) to reveal the underlying substrate as necessary.

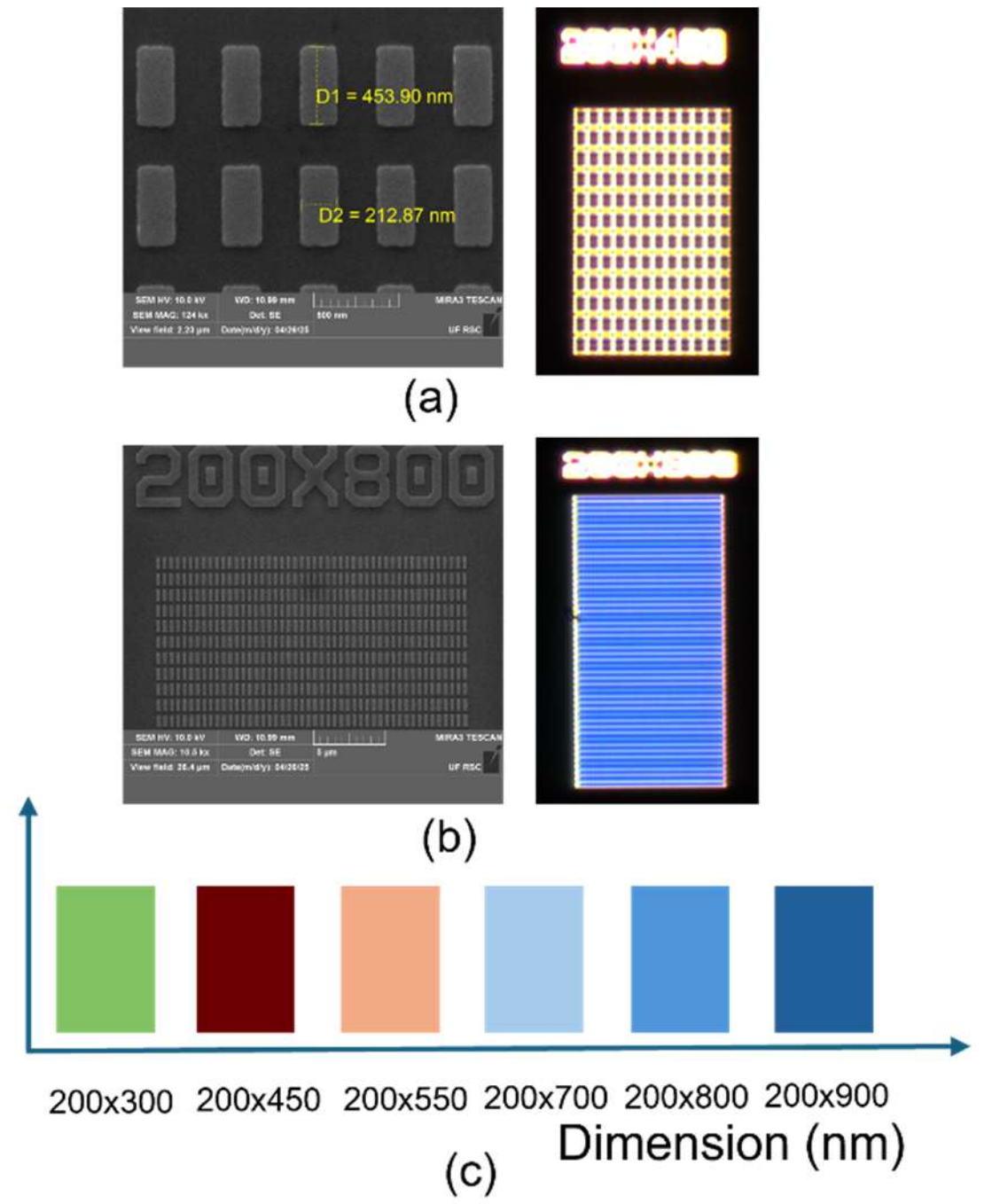

*Fig. 4: (a) & (b) Top view SEM images of Si nanopillar array and their response. (c) Optical response varies based on the nanostructures*

Scanning Electron Microscopy (SEM) was employed to inspect the fabricated structures. High-resolution SEM images confirmed that the photonic crystal patterns closely matched the intended geometries, with measured deviations within ±10 nm of the design targets. Pillars exhibited well-defined sidewalls, uniform heights (within 5 nm tolerance) as shown in Fig. 4, and minimal edge roughness. Lattice regularity and alignment were also maintained across the 20 µm × 20 µm fingerprint regions.

Optical characterization of the fabricated samples was performed using a hyperspectral microscope with a broadband white light. The reflected light spectra from each fingerprint region were collected under normal and oblique illumination, as shown in Fig. 4. The results showed distinct color patches corresponding to the resonance peaks predicted in simulations. For instance, a structure designed to resonate at ~630 nm reflected a strong red hue under normal light, while another resonating at ~520 nm appeared green.

The measured spectral peaks were consistent with FDTD predictions, with minor deviations (~5–10 nm), attributable to fabrication tolerances. These small shifts are acceptable and even beneficial, as they contribute to entropy in the fingerprinting scheme. Moreover, polarization-resolved measurements confirmed that structures designed with asymmetric pillars reflected different intensities under TE and TM illumination, reinforcing the capability of encoding additional information.

Notably, optical signatures were robust to environmental variations such as ambient lighting or mild surface contamination. Since the resonance effect is governed by structural features on the order of hundreds of nanometers, minor surface dust did not significantly affect detection. The reproducibility of optical fingerprints across multiple dies fabricated on the same wafer further demonstrated the practicality of large-scale deployment.
In conclusion, the fabrication and imaging analysis validated the core hypothesis of this work: photonic crystal nanostructures embedded in density filler regions can serve as robust, verifiable, and unique hardware fingerprints for silicon photonic devices. The combination of precise design, controlled process variability, and optical interrogation forms a solid foundation for secure authentication and supply chain assurance.

## V. Conclusion

We propose a hardware fingerprinting technique for silicon photonic integrated circuits using photonic crystal-based density-controlled patterns. This method exploits standard filler regions to embed nanostructures that produce unique, wavelength-dependent optical signatures without requiring additional fabrication steps. Finite-Difference Time-Domain (FDTD) simulations demonstrate that variations in pillar size and spacing yield distinct resonance peaks in the visible to near-infrared range. Experimental results from fabricated structures confirm the feasibility of this approach, showing strong correlation with simulated spectra and high sensitivity to design variations. The fingerprinting scheme supports multidimensional encoding through wavelength, polarization, and angle dependencies, offering strong resistance to cloning and tampering. By enabling chip-level authentication through non-invasive optical interrogation, this technique provides a scalable, low-cost, and reliable method for supply chain assurance and anti-counterfeiting in photonic systems. This approach represents a significant advancement toward secure and trustworthy silicon photonics deployment.